\documentclass[prb,preprint,aps,amsmath,amssymb,superscriptaddress]{revtex4-1}
\usepackage[T1]{fontenc}
\usepackage[utf8]{inputenc}
\usepackage[english]{babel}
\usepackage{units}
\usepackage{amsmath} 
\usepackage{amsfonts} 
\usepackage{amssymb} 
\usepackage{color}
\usepackage{graphicx}
\usepackage{epsfig}
\usepackage{epstopdf}
\usepackage{float}
\usepackage{nicefrac}
\usepackage{url}
\usepackage{amsmath}

\begin{document}

\title{Electron-phonon coupling in suspended graphene: supercollisions by ripples}

\author{Antti Laitinen}
\affiliation{Low Temperature Laboratory, Aalto University, P.O. Box 15100, FI-00076 AALTO, Finland}
\author{Mika Oksanen}
\affiliation{Low Temperature Laboratory, Aalto University, P.O. Box 15100, FI-00076 AALTO, Finland}
\author{Aur\'elien Fay}
\affiliation{Low Temperature Laboratory, Aalto University, P.O. Box 15100, FI-00076 AALTO, Finland}
\author{Daniel Cox}
\affiliation{Low Temperature Laboratory, Aalto University, P.O. Box 15100, FI-00076 AALTO, Finland}
\author{Matti Tomi}
\affiliation{Low Temperature Laboratory, Aalto University, P.O. Box 15100, FI-00076 AALTO, Finland}
\author{Pauli Virtanen}
\affiliation{Low Temperature Laboratory, Aalto University, P.O. Box 15100, FI-00076 AALTO, Finland}
\author{Pertti J. Hakonen} \email{pertti.hakonen@aalto.fi}{}
\affiliation {Low Temperature Laboratory, Aalto University, P.O. Box 15100, FI-00076 AALTO, Finland}
%

%
\begin{abstract}

Using electrical transport experiments and shot noise thermometry, we find strong evidence that "supercollision" scattering processes by flexural modes are the dominant electron-phonon energy transfer mechanism in high-quality, suspended graphene around room temperature. The power law dependence of the electron-phonon coupling changes from cubic to quintic with temperature. The  change of the temperature exponent by two is reflected in the quadratic dependence on chemical potential, which is an inherent feature of two-phonon quantum processes.
\end{abstract}

\maketitle
%


 Electron-phonon coupling is the basic means to
 control energy transport in a variety of devices which provide extreme sensitivity in calorimetry, bolometry, and
radiation detection (infrared/THz). Owing to thermal noise, such devices are typically operated at cryogenic temperatures, at which the coupling between electrons and phonons becomes weak. Under these conditions, graphene is expected to have an advantage owing to its small heat capacity that allows fast operation even though its electron-phonon coupling becomes exceedingly small near the Dirac point \cite{Kubakaddi2009,Tse2009,Bistritzer2009,Viljas2010,Betz2012,Suzuura2002,Katsnelson2012}.

The weak electron-acoustic phonon coupling in graphene stems from the restrictions in energy transfer caused by momentum conservation \cite{Katsnelson2012}. The maximum change of momentum at the Fermi level is twice the Fermi momentum $2 k_F$, which corresponds to phonon energy $\hbar \omega_{2k_F}$. This energy defines a characteristic temperature, the Bloch-Gr\"uneisen temperature by $\hbar \omega_{2k_F}= k_B T_{BG}$, above which only a fraction of phonons are available for scattering with electrons in the thermal window. Manifestations of this has been observed in resistance vs. temperature measurements \cite{Chen2008}, especially on electrolytically-gated graphene \cite{Efetov2010}.

The energy transfer limits due to momentum conservation are circumvented when acoustic phonon scattering is assisted by other scattering processes, a.k.a. supercollision events, or when energy is transferred to flexural phonons, which is a two-phonon process. \cite{Song2012,Graham2012,Betz2012a,Graham2013,WeiChen2012,Mariani2010,Castro2010,Katsnelson2012} The flexural phonon mechanism can dominate over the single acoustic phonon process, but to win over supercollision cooling, extremely clean samples are required. \cite{Song2012} Moreover, even if a sample is free of impurities, other scattering mechanisms such as static or dynamic ripples can still facilitate supercollisions.

The heat flow from the conduction electrons/holes to the lattice can be characterized by a power law of the form $P=\Sigma (T_e^{\delta} -T_{ph}^{\delta})$, where $T_e$ is the electron temperature, $T_{ph}$ the phonon temperature, $\Sigma$ the coupling constant and $\delta $ a characteristic exponent \cite{Giazotto2006}.  Our monolayer graphene experiments have been conducted near the Dirac point at charge densities $n < 0.1-4.5 \cdot 10^{11}$ cm$^{-2}$, which corresponds to $T_{BG} < 42$ K for longitudinal acoustic phonons (and even lower for transverse acoustic and flexural phonons). Consequently, all our results have been measured in the regime of $T > T_{BG}$, where the scattering of electrons from acoustic phonons leads to $\delta = 1$ or $\delta = 5$, depending whether $T_e << \mu$ or $T_e > \mu$, respectively \cite{Viljas2010}. Non-conventional pathways for cooling, the supercollision cooling or flexural two phonon processes \cite{Song2012}, yield $\delta = 3$ or $\delta = 5$ for $T_e < \mu$ or $T_e > \mu$, respectively.

In this work, we have employed shot noise thermometry in combination with conductance measurements to determine the electron-phonon coupling in high-quality, suspended graphene monolayers. We demonstrate that cooling in our graphene samples, even though their field effect mobility is $\mu_f > 10^5$ cm$^2$/Vs at low temperatures, takes place via supercollision phonon processes at temperatures $T_e = 200-600$ K. We observe a power-law dependence having $\delta \simeq 3 - 5$, depending on the magnitude of the chemical potential $\mu = 10 - 73$ meV relative to $k_B T_e$. The change in the exponent agrees well with the cross-over behavior between $\delta = 5$ and  $\delta = 3$ for the non-degenerate ($k_BT_e>\mu$) and degenerate cases ($k_BT_e<\mu$), which also leads to a $\mu^2$ dependence seen in the data as a  function of chemical potential. We also find that the location of the cross-over between the two regimes is shifted by the Fermi velocity renormalization caused by interactions. For the deformation potential, we obtain $D \simeq 64$ eV over the full range of data. No increased relaxation rate due to optical phonon scattering was found in our experiments at $T_e < 600$ Kelvin. The origin of the supercollision scattering can be understood to be small-scale ripples in the suspended graphene.

\begin{figure}[H]
\centering
\includegraphics[width=0.8\textwidth]{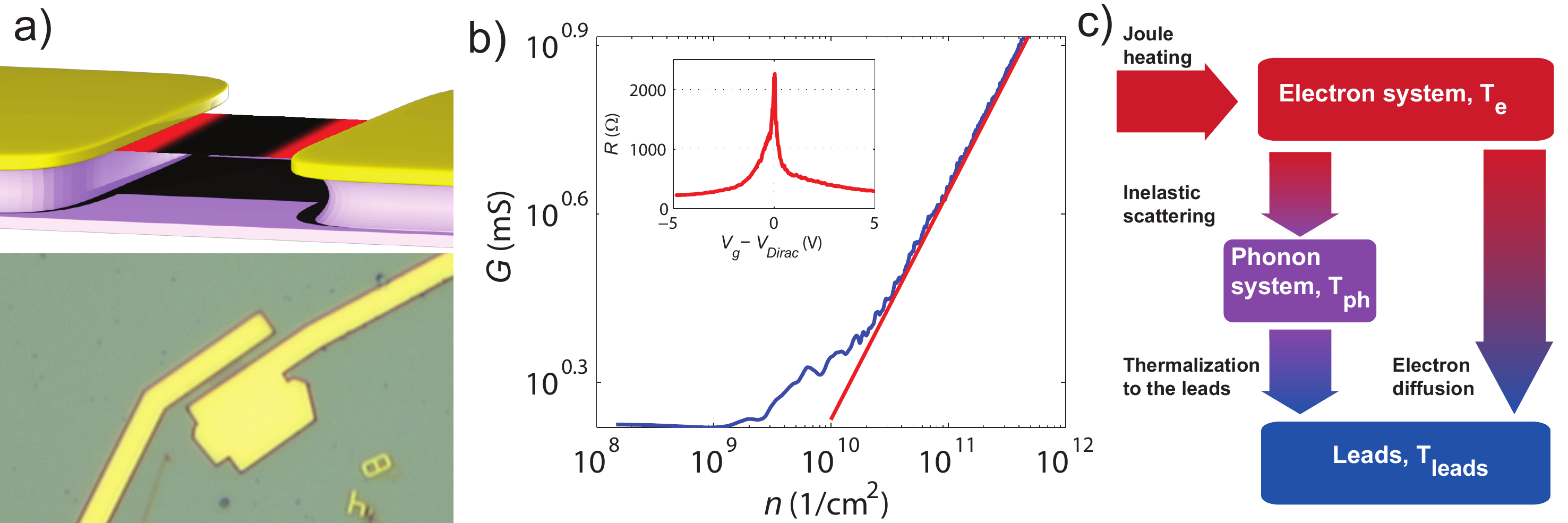}
\caption{a) Schematic view of a suspended graphene sample, where metallic leads contact a graphene flake (top part), and an optical image of the sample (lower part). A layer of silicon oxide, part of which is etched away, separates the circuit from the back gate made of heavily doped silicon.   b) Logarithm of the conductivity as a function of the logarithmic charge carrier density $ n=C_g (V_g-V_g^D)/e$. The solid line extrapolates towards the residual charge carrier density of $n_0=0.8 \cdot 10^{10}$ cm$^{-2}$; the offset of the Dirac point  due to charge doping is corrected. The inset displays zero-bias resistance versus  $V_g$.  c) Thermal model of the sample: electrons are thermalized to metallic leads via heat diffusion and electron-phonon coupling conduction in series with Kapitza boundary resistance to the leads/substrate phonons.}
\label{fig:schema}
\end{figure}

The studied sample, with length $L=1.1$ $\mu$m and width $W=4.5$ $\mu$m, was manufactured from exfoliated graphene using e-beam lithography and etching in hydrofluoric acid. Raman spectroscopy was employed to verify that the sample was a monolayer graphene sheet. Before measurements, the graphene was annealed by passing a current of 1.1 mA through the sample (nearly 0.9 V in voltage) which evaporated resist residuals and resulted in a nearly neutral, high-quality sample with the charge neutrality (Dirac) point located at $V_g^{D} = +0.4$ V.   The gate capacitance was determined from the parallel model (and verified by the $\mu$ scale in Fabry-P\'erot interference patterns \cite{Oksanen2013}): $C_g$ = 4.7 $ \frac{nF}{cm^2}$, which is  close to 6 $\frac{nF}{cm^2}$ in a similar device reported in Ref. \onlinecite{Bolotin2008b}.
The residual charge carrier density was determined to be $n_0= 0.8 \cdot 10^{10}$ $\frac{1}{cm^2}$ from $\log(G)$ vs $\log(n)$ plot (see Fig.  \ref{fig:schema}b) as a point where the linear behaviour levels to a constant value at low charge densities $ n=C_g (V_g-V_g^D)$. Note that the slope of $G(n)$ on the log-log plot is nearly 1/2 which indicates ballistic behavior. The effective $n_0$ is 20\% larger when making a similar log-log plot for the electron-phonon coupling vs. $n$.
The initial slope of $G(n)$ was employed to determine the field effect mobility of $\mu_f > 10^5$ cm$^2$/Vs. The inset of Fig. \ref{fig:schema}b displays the variation of zero-bias resistance $R_0=dV/dI_{|V=0}$ over chemical potentials ranging $\pm 73$ meV across the Dirac point. The measured minimum conductivity falls below $\frac{4e^2}{h}$ approaching theoretical minimum conductivity $\sigma _{min} = \frac{4e^2}{\pi h}$ for high aspect ratio samples \cite{katsnelson2006a,Tworzydlo2006}; the actual maximum resistance of the sample was 2.2 k$\Omega$.

In addition to \textit{IV}-characteristics and differential conductance $G_d = \frac{dI}{dV}$, we measured zero-frequency shot noise $S_I$ and $dS/dI$ (over 600-900 MHz) calibrated against a AlO$_x$ tunnel junction in the same cooldown (see Ref. \onlinecite{danneau2008b} for details). Poissonian noise is given by $S_p = 2q\langle I\rangle$ where $\langle I\rangle$ is the average current. The Fano factor defines the noise level $S_I$ compared to the Poissonian noise \cite{Blanter2000}, $F = S_I /S_p$. In addition to gate voltage, the Fano factor depends also on the bias voltage $V$. When the bias voltage is increased, electron-electron interaction effects (in the so called "hot electron regime") and inelastic scattering effects start to influence the Fano factor. For diffusive transport in local equilibrium, the Fano factor is related to electronic temperature: \cite{wu2010,santavicca2010} $T_e = \frac{Fe|V|}{2k_B}$ . This equation is the basis of our noise thermometry on graphene electrons. Our experiments were performed on a dilution refrigerator operated typically around 0.5 K in our large bias experiments. For experimental details, we refer to Ref. \onlinecite{Oksanen2013}.

%

\begin{figure}[h!]
\centering
\includegraphics[width=0.95\textwidth]{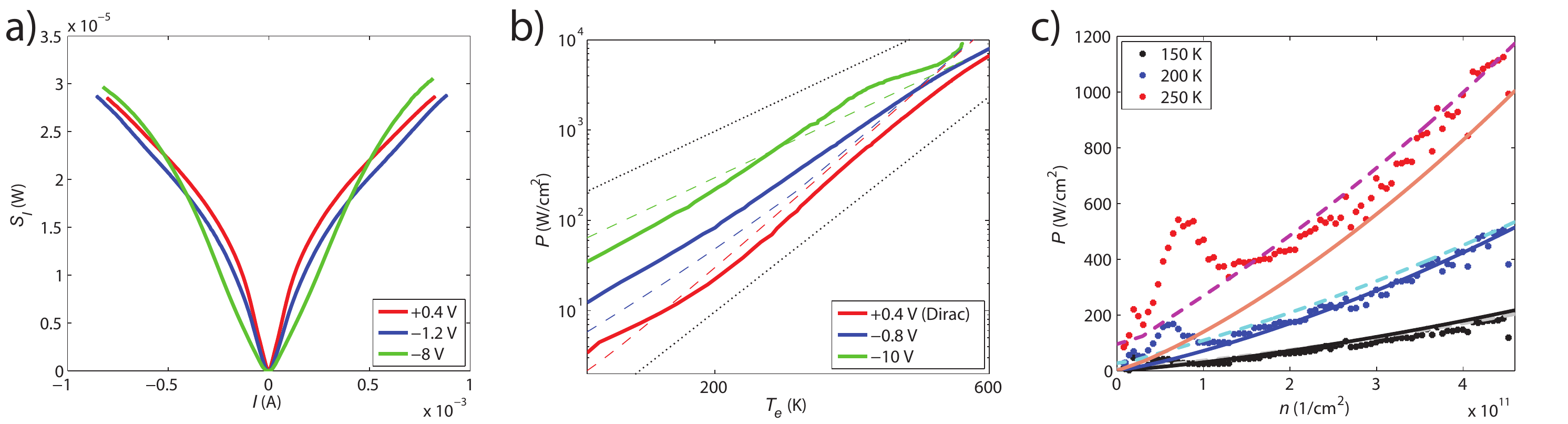}
\caption{a) Shot noise vs. the bias current $I$ at a few values of gate voltage near the Dirac point at $V_g=+0.4$ V ($V_g$ is given in the figure).
 b) Heat flow to phonons $P$ as a function electron temperature $T_e=Fe|V|/2k_B$.
 The  green, blue, and red curves represent measured data while the dashed lines display the theoretical behavior from Eq. \ref{theory} with $k_F \ell =3$ and $D=64$ eV; for other parameters, see text. The dotted lines denote power laws with exponents $\delta=3$ (upper) and $\delta =5$ (lower). c) Electron-phonon coupling power as a function of the gate-induced charge carrier density $n$ at temperature $T_e$ = 150, 200,  and 250 K for $V_g <V_g^D=+0.4 $ V (no \textit{pn}-junction). The  theoretical lines are obtained from Eq.  \ref{theory} using the same parameters as in  Fig. \ref{fig:results}b: the lines denote the expansions at $\mu << k_B T_e$ (dashed lines) and  at $\mu >> k_B T_e$ (solid lines).
}
\label{fig:results}
\end{figure}

 Measured shot noise $S_I$ up to high bias is illustrated in Fig. \ref{fig:results}a. Near the Dirac point, strong initial increase in $S_I$ is found, which reflects a clearly larger Fano factor at small charge densities $n_0$ than at $n >> n_0$. Rather smooth, symmetric increase in $S_I$ is observed with bias current over the full range of bias conditions.  In our experimentally determined e-ph coupling, there is a difference in the prefactor of the power laws obtained at equal hole and electron densities. We assign this to the presence of $pn$-junctions at $V_g > V_g^D$, which undermines the accuracy of our shot noise thermometry: at large bias voltages with enhanced phonon scattering, the noise generated by the $pn$-junctions is an additive quantity to the local thermal noise, and the resulting electronic temperatures are higher than what they really should be. Consequently, we rely only on the data without \textit{pn}-junction in our actual analysis. The shot noise temperature is corrected for contact resistance, which increases the measured temperatures by 20\% in overall.

The Joule heating power to graphene electrons equals to $P_e = (V- \Delta V_{cc}) I$ where $\Delta V_{cc} $ denotes the total voltage drop over the contacts $\Delta V_{cc} = R_c I $, where $R_c =40$ $\Omega$.
The heat flow paths in the sample, namely diffusive transport and inelastic scattering, that balance $P_e$ are depicted in Fig. \ref{fig:schema} c.  When the graphene lattice is at liquid helium temperatures or below and the electron system is heated to $200-300$ K, the electronic heat will flow  mostly to acoustic/flexural phonons at rate $P=\dot{Q}_{e-ph} = \Sigma \left(T_e^\delta -T_{ph}^\delta \right)$, not to contacts by electronic diffusion. Under this Joule heat flux, elevation of the graphene acoustic phonon temperature should remain below a few tens of Kelvins \cite{Balandin2011}, and the latter term $T_{ph}^\delta $ can be neglected. In order to improve the precision in the determination of $\delta$, we subtract the electronic heat conduction to obtain $P$ from $P_e$.
At $T_e < 100$ K, we observe power-law behavior in $P_e$ vs $T_e$ with an exponent $\delta \simeq 1.6$ \cite{Tarkiainen2003}, but our accuracy is not enough to test modifications of the Wiedemann-Franz law as achieved in Ref. \cite{Fong2013}.

As indicated by the data on $P$ vs $T_e$ in Fig. \ref{fig:results}b, we observe for the electron-phonon heat flow a power-law dependence having $\delta \simeq 3 - 5$, depending on the chemical potential $\mu = 10 - 73$ meV:
we find $\delta=5$ at $|V_g-V_g^{D}| < 0.6$ V while, at $|V_g-V_g^{D}| > 5$ V, we observe $\delta=3$. At $0.6$ V$< |V_g-V_g^{D}| < 5$ V, our data display a cross-over from $\delta=5$ to $\delta=3$. This cross-over takes place when the chemical potential reaches $\simeq 20$ meV, i.e. when the electronic system starts to become degenerate ($k_B T_e/\mu <1$).  This finding $\delta = 5$ at $\mu << k_BT_e$ is consistent with single acoustic phonon scattering at high temperatures \cite{Kubakaddi2009,Viljas2010}, but the theoretical expectation for acoustic phonons at $\mu >> k_BT_e$ disagrees with our results.  Furthermore,  the magnitude of the predicted e-ph coupling for  $\delta=5$ is too small unless an unrealistically large value for the deformation potential is adopted. Moreover, instead of the expected $\mu$-dependence of acoustic phonon scattering ($P\propto{}\mu^4$ at $T_{BG}<T<\mu$; $P\propto\mu^0$ at $T_{BG},\mu<T$), we find $\mu^2$-dependence (or more precisely, linear in $n$, see below) which contradicts single phonon scattering but agrees with flexural phonon and supercollision scattering. Hence, we may rule out single phonon scattering events  \cite{Kubakaddi2009,Viljas2010} as the dominant e-ph coupling in our suspended sample.

The electron-phonon energy flow via supercollisions is predicted in Ref. \onlinecite{Song2012}. Taking into account the energy dependence of the density of states in graphene, which leads to the cross-over between the $T_e^3$ and $T_e^5$ power laws, the result reads:
\begin{equation}\label{theory}
  \dot{Q}_S
  \rvert_{T_{\rm ph}=0}
  \simeq
  -
  \frac{N}{(2\pi\hbar^2v_{F}^2)^2}
  \frac{g_D^2 k_B^5}{\hbar k_F \ell}
  T_e^5
  q(\frac{\hbar v_{F}\sqrt{\pi n}}{k_BT_e})
  \,,
\end{equation}
where $q(z) \simeq 9.69 + 1.93 z^2$ for $|z|\ll 5$ and $q(z) \simeq 2\zeta(3)z^2 \approx 2.40 z^2$ for $|z|\gg 5$, and $N=4$ is the degeneracy in graphene. The electron-phonon coupling constant for deformation potential is $g_D=D/\sqrt{2\rho s^2}$ where $D$ is the deformation potential and $\rho$ the mass density of the graphene sheet. The quantity $k_F\ell$ is the dimensionless mean free path associated with the additional scattering mechanism enabling the supercollisions. For experimental convenience, we have also here written the above in terms of the particle density $n$ which is kept fixed by gate voltage. We take renormalization of the Fermi velocity by electron-electron interactions \cite{Kotov2012,Elias2011} into account by setting $v_F=v_F(n,T_e)$ with dependence on both charge density and electronic temperature as detailed in the supplementary material.

Supercollision and flexural phonon scattering yield similar power law predictions \cite{Song2012}, but can be distinguished from each other by comparing numerical estimations against measured data. Although flexural phonon scattering may dominate over single acoustic phonon scattering (when $T_e > 10 T_{BG}$, where $T_{BG}$ refers to acoustic phonons), the ratio of heat flux via flexural modes $\dot{Q}_F$ to that of supercollision heat flow $\dot{Q}_S$ is estimated as $\dot{Q}_F/\dot{Q}_S = k_F \ell/200$.
We may fit the supercollision formula Eq. \ref{theory} to our data $\propto T_e^3$ at  $|V_g-V_g^{D}| > 5$ V using the parameters $v_{F}= 1.0 \cdot 10^6$ m/s (at high energy) and  $D/({k_F}\ell)^{1/2}=37$ eV. Using the the  theoretical estimate $D=20 - 30$  eV  \cite{Suzuura2002,Katsnelson2012}, we may conclude that only supercollision cooling is compatible with the results, as the heat flux is 100 times larger than what can be produced by flexural two-phonon scattering.

Fig. \ref{fig:results}c displays the charge carrier density dependence of the electron-phonon cooling power $P$ at a fixed temperature. The calculated curves indicate the variation on $n$  obtained from the low- and high-$\mu$ expansions of Eq. \ref{theory} using the same parameters as above.
Comparison of the data at 150, 200, and 250 K indicate that the data scale with the ratio $k_B T_e/\mu$ in accordance with the theory. The evident bumps at low/intermediate particle density are assigned to resistance fluctuations that may cause bias-dependent modification of noise even at the band 600-900 MHz \cite{Wu2007}. The linear slope in $n$  corresponds to $\mu^2$ dependence away from the Dirac point where the velocity renormalization is negligible.

The value for ${k_F}\ell$ for supercollision scattering is small for our sample at high bias. Possible mechanisms contributing to this, in addition to short-range impurities, are dynamic and static ripples \cite{Zan2012}, as well as scattering from the edges (possibly magnetic) of the sample. Coulomb impurities are not expected to contribute significantly to supercollisions. Ripples can yield quite small ${k_F}\ell$ for scattering, as estimated in Ref.~\cite{song_v1}; STM results of Ref. \cite{Zan2012} indicate ${k_F}\ell \simeq 2$ for a typical current-cleaned graphene sample such as ours at room temperature (${k_F}\ell=1.8$ using amplitude of corrugation $Z=0.4$ nm and radius of curvature $R=1.5$ nm, i.e. wave length $\sim 5$ nm \cite{Zan2012}). Both static and dynamic ripples are known to influence the resistance of suspended graphene samples \cite{Mariani2010,Castro2010}.

Further information of the origin of the short $k_F\ell$ in our experiment is found from the results of resistance measurements in the same sample. At low bias, the results indicate nearly ballistic behavior, which rules out scattering from static frozen-in ripples or short-ranged impurities. Using our total square resistance $R_{\square}=(V/I)W/L$ vs. $T_e$ curve at $n \simeq 0.8 \cdot 10^{11}$ cm$^{-2}$ (see the supplementary information), we find the temperature-dependent component in the resistivity that increases as $T_e^2$ having a magnitude of $\rho_i=0.01 (T_e/\textrm{K})^2$. This value for $\rho_i $  is quite close to the results of Ref. \cite{Castro2010}  on flexural phonon limited mobility in suspended graphene. By interpreting $\rho_i $ directly as a scattering length via $k_f\ell_r=h/(4e^2\rho_i)$, we find a small value $k_F\ell_r\sim7$ at $T_e =300$ K, indicating
an effective scattering mechanism activated by the large bias; the subscript $r$ refers to scattering events governing the resistance.

The effect of quasielastic scattering from thermally excited flexular phonons on the supercollision energy flow is analyzed in Ref. \onlinecite{Virtanen2013}. An effective $k_F\ell_{\rm eff}$ for supercollisions can be estimated from the resistance, given an adjustment for differences in characteristic wave vector scales. Scattering leading to resistance occurs mainly on scales of $\max(k_F, q_*)$, where $q_*= \sqrt{u} s/\kappa$ denotes the cut-off of the flexural modes due to strain $u$ in the membrane. \cite{Castro2010} Supercollision energy transport is dominated by thermal phonons with $q_T  = k_B T/\hbar s$. Here, $s=2.1 \cdot 10^4$ m/s is the speed of the acoustic mode in graphene, and $\kappa=4.6\cdot10^{-7}$ m$^2$/s is specified by the dispersion relation of the flexural modes $\omega = \kappa q^2$.
When $k_F < q_* < q_T$ (regime VI of Ref. \onlinecite{Mariani2010}), this leads to an effective value  $\left[{k_F}\ell_E\right]^{-1} \simeq 10(q_*/q_T)^2 (4e^2/h)\rho_i{\cal G}$, where $\rho_i$ $(\propto\left[{k_F}\ell_r\right]^{-1})$ denotes the $T_e^2$ dependent part of the resistivity and ${\cal G} \propto \log (q_*/q_T)$ is a factor of the order of unity; the subscript $E$ denotes that these scattering events govern the energy exchange.
We have estimated the strain $u\approx 4-8 \cdot 10^{-4}$  from the gate voltage dependence of the measured fundamental flexural mode frequency $f_{res} = 77$ MHz: $|f-f_{res}| < 0.1$ MHz over $V_g = 0 - 10$ V. In our analysis we employ $u = 4 \cdot 10^{-4}$, which minimizes the difference between our measured $\rho_i$ and the calculated result $\rho_i \propto \frac{1}{u}$. Combining $u = 4 \cdot 10^{-4}$ with the value for $\rho_i$ yields a temperature-independent result ${k_F}\ell_E \simeq 3$ that can be employed in Eq. \ref{theory}; this bias-induced value for ${k_F}\ell_E$ is much smaller than expected for short-range scatterers in high-quality graphene.
However, smaller strain and $q_*$ and hence larger $k_F\ell_E$ would be necessary for the results to be in line with previous results on resistance due to flexural phonons, although these results are not fully comparable as they were measured at small bias voltage. \cite{Castro2010}

Taking ${k_F}\ell$ as a materials parameter independent of the variation in the charge carrier density, as is the case above, and using the above parameters in Eq. \ref{theory}, we are able to fit our results also in the non-degenerate case $\propto T_e^5$ and in the cross-over regime as depicted in Fig. 2b.  The fit indicates $D=64$ eV which is by a factor of two larger than the theoretical estimate $D=20-30$ eV for the unscreened case \cite{Suzuura2002,Katsnelson2012}.  Altogether, once the values of $D = 64$ eV and ${k_F}\ell=3$ are fixed, the overall features of our data are accounted for by Eq.  \ref{theory}.

Compared with previous transport experiments \cite{Betz2012,Betz2012a}, our results on the electron-phonon coupling are rather close in magnitude in the degenerate limit at high bias, even though our samples are ballistic at low bias. This suggests that the mechanism for the electron-phonon scattering may be similar, i.e. it could be ripples in both experiments. A strong contribution is expected to arise from excited flexural modes, since they can be in a "hot-phonon-like" non-equilibrium state as has been observed with optical phonons in carbon nanotubes under similar conditions \cite{Dekker2000,Mauri}. Such non-equilibrium state of ripples, however, is more likely in a suspended sample than in a supported device.

Finally, photocurrent in suspended graphene p-n junctions has been found to be by an order of magnitude larger than in supported structures; this enhancement has been attributed to the elimination of a dominant electronic cooling channel via the surface phonons of the polar SiO$_2$ substrate \cite{Freitag2013}. Similar direct coupling to phonons has also been observed in single walled carbon nanotubes \cite{Baloch2012}. Our experiments avoid these problems and address truly the inherent electron-phonon coupling in graphene.

In summary, our experiments indicate strong supercollision cooling in the presence of ripples in suspended graphene. We have achieved the first results on graphene cooling in the high temperature limit demonstrating the $T_e^5$ dependence for the electron-phonon heat transfer. In the low-$T_e$ limit, our results indicate quadratic dependence on the chemical potential, which is a characteristic signature of non-conventional cooling processes. This $\mu^2$ behavior is in line with the cross-over, found at $T_e \simeq \mu/k_B$, from the quintic high-$T_e$ behavior to cubic in the low-$T_e$ regime. Our analysis yields for the deformation potential $D=64$ eV.  Most likely, the observed strong electron-phonon coupling originates from supercollision events that are facilitated by a high-bias-induced ripple structure with a magnitude in line with recent room temperature STM experiments on suspended graphene.

We acknowledge fruitful discussions with T. Heikkil\"{a} and M. Katsnelson. We have benefited from interaction with M.F. Craciun and S. Russo within a scientific exchange programme between Low Temperature Laboratory and Centre for Graphene Science at Exeter University.  Our work was supported by the Academy of Finland (contracts no. 135908 and 250280, LTQ CoE), by the EU-project RODIN FP7-246026, and by the European Science Foundation (ESF) under the EUROCORES Programme EuroGRAPHENE. This research project made use of the Aalto University Cryohall infrastructure. MO is grateful to V\"{a}isäl\"{a} Foundation of the Finnish Academy of Science and Letters for a scholarship.


\providecommand*\mcitethebibliography{\thebibliography}
\csname @ifundefined\endcsname{endmcitethebibliography}
  {\let\endmcitethebibliography\endthebibliography}{}

\newpage
%
%
%

\section*{Supplementary information for \\
"Electron-phonon coupling in suspended graphene: supercollisions by ripples"}

%

%
%

\section{Renormalization of the Fermi velocity }

In suspended graphene samples close to the Dirac point, the electronic spectrum and the density of states is renormalized by electron-electron interactions, as discussed in Ref. \onlinecite{Kotov2012}. Effectively, the Fermi velocity increases at low energies, which then is also reflected in the supercollision mechanism.
Here, we account for the interaction effects via a simple renormalization group (RG) procedure~\cite{Kotov2012}. As electron states $|\vec{k}|>\Lambda$ are integrated out, the effective electron-electron interaction $\alpha=e^2/(4\pi\epsilon_0\hbar v_F)$ flows towards zero. The corresponding velocity renormalization in the simplest approximation reads
\begin{align}
  \label{eq:effective-dispersion}
  \tag{S1}
  v_F(\vec{k})
  \sim
  v_{F,*}
  \Bigl(
  1
  +
  \frac{\alpha_*}{4}
  \ln\frac{
    \Lambda_*
  }{
    \max(|\vec{k}|, \sqrt{\pi n}, k_BT_e/v_{F,*})
  }
  \Bigr)
  \,,
\end{align}
where the renormalized velocity at scale $\Lambda_*$ is set as $v_{F,*}=1.0 \cdot 10^6$ m/s according to regular tight binding parameters \cite{Katsnelson2012}, $  \alpha_* = \frac{\alpha}{\epsilon_G} =  1$ with  $\epsilon_G = 2.2$,\cite{Elias2011} and the scale $\Lambda_* $ is calculated at the charge density  $\unit[5\times10^{12}]{cm^{-2}}$ yielding $\Lambda_* =  \frac{k_B}{\hbar v_{F,*}} \times \unit[2200]{K}$.
The Fermi velocity is momentum-dependent and renormalized at $\mu=0$ to $v_F(|\vec{k}|)$ at $|\vec{k}|<\Lambda_*$. Temperature and chemical potential both cut off the flow of $v_F$, although the cross-over region when $\hbar k_F \sim k_B T_e/ v_{F,*}$ is not completely accurately handled by this equation.

Electron-phonon supercollisions involve, in addition to initial and final low-energy states whose energies are renormalized as described by Eq.~\eqref{eq:effective-dispersion}, a high-energy virtual electron state at
$|\vec{k}|\sim|\vec{q}|\sim{}k_BT_e/(\hbar s)\gg{}k_F$; here $s$ stands for the speed of sound.
For typical temperatures in our case, this is close to or above the cutoff $\Lambda_*$, where the variation of $v_F$ is small, and we take $v_F\approx{}v_{F,*}$ in this regime. As the velocity renormalization results to a linear spectrum below the Fermi level, the $\mu$ vs. $n$ relationship remains simple, $\mu=\hbar{}v_F(n,T_e)\sqrt{\pi{}n}$. With these two provisions, the effect of velocity renormalization on supercollisions is obtained by substituting Eq.~\eqref{eq:effective-dispersion} for $k=k_F$ into Eq.~(1) of the main text.

The main effect of the electron-electron interactions is to shift the cross-over point between degenerate and non-degenerate regimes towards smaller charge densities as compared to the noninteracting situation.  For $n\sim\unit[10^{10}]{cm^{-2}}$, we have
$v_F/v_{F,*}\sim1.3\ldots{}1.6$ between $T=\unit[10\ldots500]{K}$.

\section{Details of the electrical characteristics}
\begin{figure}[H]
\centering
\includegraphics[width=0.75\textwidth]{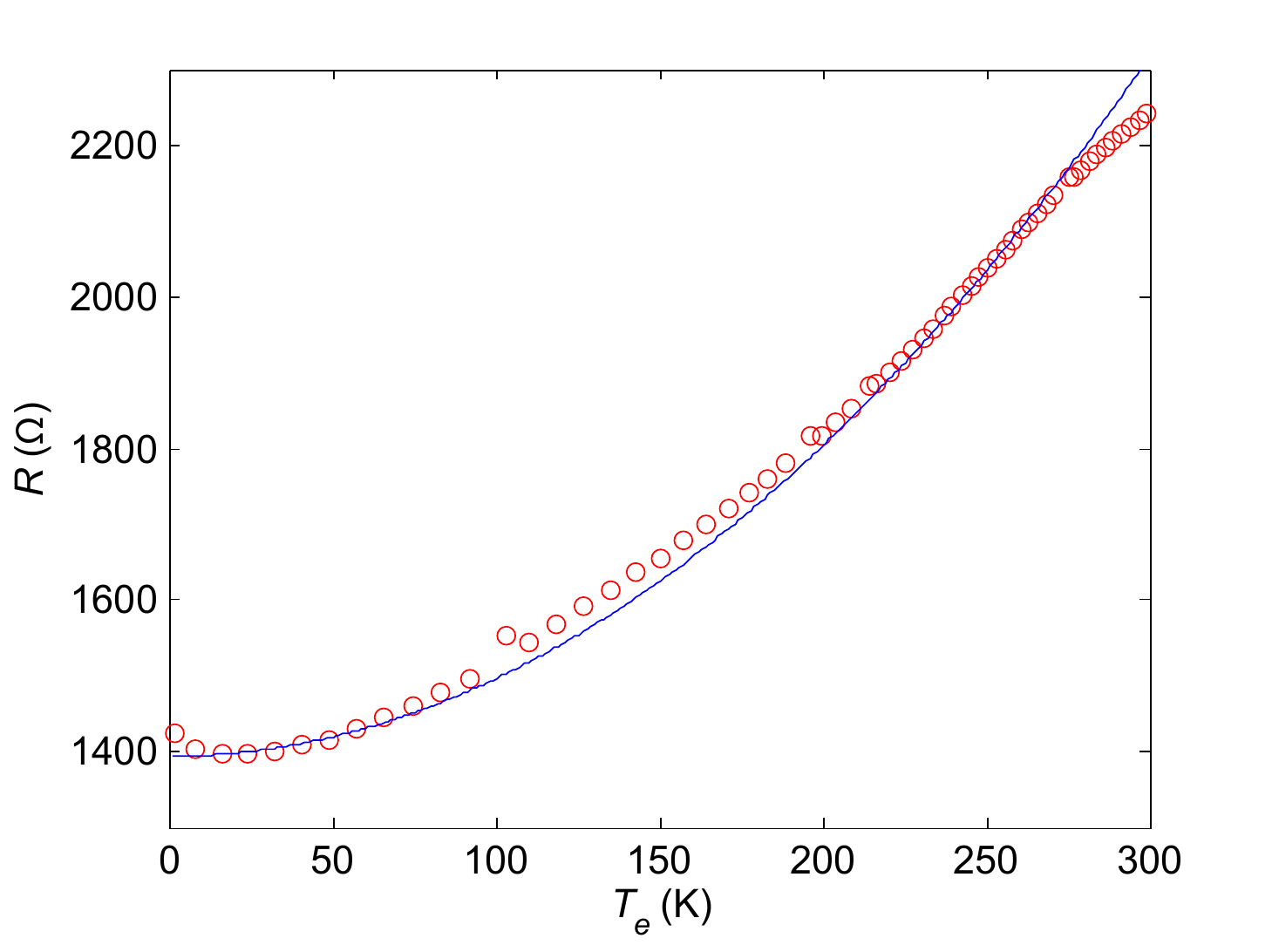}
\caption{Total square resistance $R_{\square}=(V/I) W/L$ as a function of the electronic temperature deduced from the Fano factor using $T_e=Fe|V|/2k_B$ ($n = 0.8 \cdot 10^{11}$ cm$^{-2}$). The fit function is $R_{\square}=[1390 + 0.01 (T_e/\textrm{K})^2]$  $\Omega$.}
\label{fig:RvsT}
\end{figure}

In our experiments, we measured the electrical characteristics of the sample both at DC and at low-frequency AC (dynamic resistance).
The strength of the flexural phonon scattering was determined from the measured total square resistance $R_{\square}=(V/I) W/L$ illustrated in  Fig.  \ref{fig:RvsT}  at $0.8\cdot10^{11}$ cm$^2$ ($V_g$ = -2.4 V) as a function of $T_e$, with $ T_e$ determined from the shot noise thermometry. The behavior of $R_{\square}$ is well fit with a quadratic temperature dependence as expected for flexural phonon scattering \cite{Castro2010}. From the fit we obtain $R_{\square}=[1390 + 0.01 (T_e/\textrm{K})^2$] $\Omega$ which was employed for determining the pure temperature-dependent part $\rho_i=0.01 (T_e/\textrm{K})^2$ employed in the main text.

We also measured the zero-bias AC resistance $R_0$ as a function of temperature which turned out to be irregular because residual gas got desorbed from surfaces in the vacuum chamber while the cryostat was warming up. The desorbed gas became partly readsorbed on to the clean, current-annealed sample, causing a shift of the Dirac point due to charge doping, most likely due to oxygen. Nevertheless, the data could be fit pretty well with quadratic temperature dependence, yielding a $T_e^2$-term close to the above $\rho_i$. However, we consider the determination based on $V/I$ more reliable than our $R_0$ analysis, because it probes the scattering under the same conditions as those prevailing  in the actual electron-phonon coupling measurement.

\begin{figure}[H]
\centering
\includegraphics[width=0.75\textwidth]{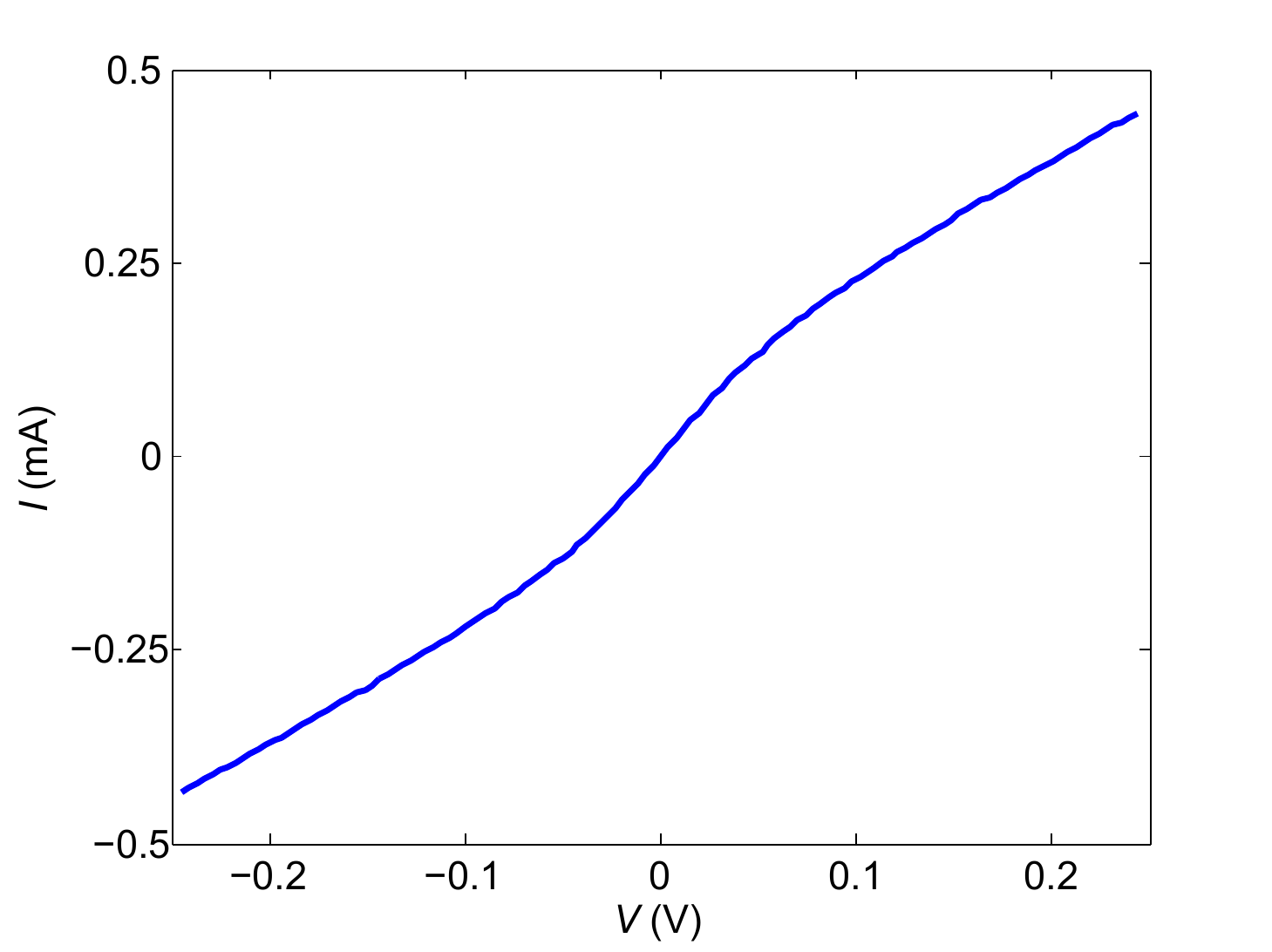}
\caption{IV-curve at $V_g$ = -2.4 V measured over a voltage range in which the electronic temperature $T_e$, determined from the shot noise thermometry, traverses the same $T-$range as displayed in  Fig. \ref{fig:RvsT}.}
\label{fig:IV}
\end{figure}

The IV-curve measured at the same charge density $n = 0.8 \cdot 10^{11}$ cm$^{-2}$  is illustrated in Fig. \ref{fig:IV}. The IV-curve reflects the clear increase of $R_{\square}$ along with the bias voltage $V$. The differential resistance $R_d= dV/dI$, measured by lock-in methods, corresponds to the inverse slope of the IV-curve. The main purpose of $ R_d(V )$ measurements was to determine the coupling strength of the noise from the sample to the preamplifier.

\section{Analysis of the power laws}
\begin{figure}[tb]
\centering
\includegraphics[width=0.75\textwidth]{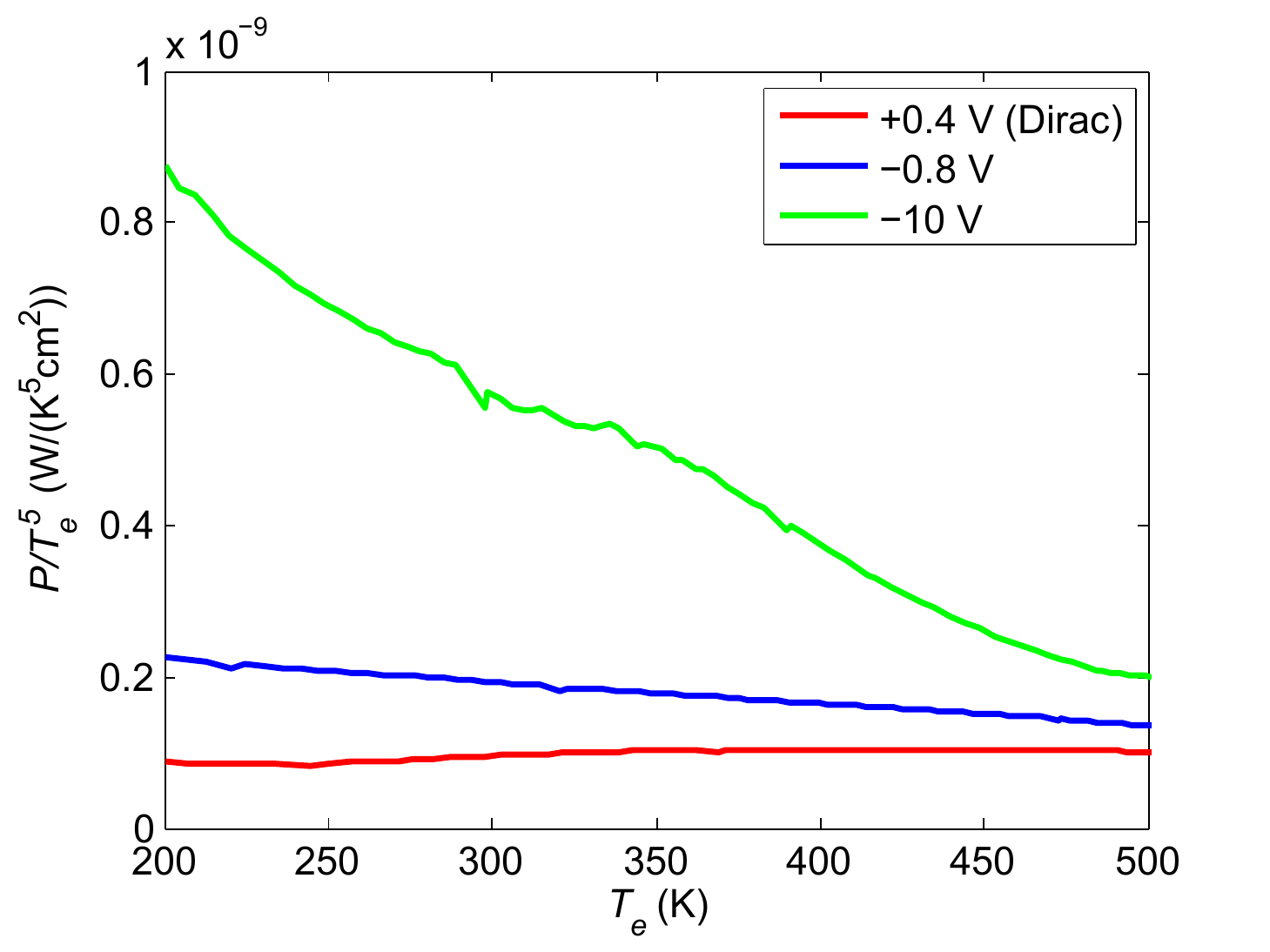}
\caption{Measured heat flow from electrons to phonons (data from Fig. 2c in the main text) normalized by $T_e^5$ and displayed as a function of $T_e$ for three gate voltage values indicated in the figure.}
\label{fig:PvsT5}
\end{figure}

The heat flow from electrons to the phonon system is characterized by the power law $P \propto T_e^{\delta}$ (where we have dropped the small  phonon temperature term $T_{ph}^{\delta}$). We observe power laws $\delta$ = 5 and 3, near and far from the Dirac point, respectively. This transition was illustrated in the main text using direct plotting on log-log frame. Here we replot the data in Fig. \ref{fig:PvsT5} by normalizing the power flow $P$ with $T_e^5$.
The curve closest to the Dirac point is almost completely flat, which indicates $T_e^5$-dependence. Slightly off from the Dirac point ($V_g$ = -0.8 V) we observe a weakly declining curve, which belongs to the cross-over region from $\delta$ = 5 towards $\delta$ = 3. Far away from the Dirac point ($V_g$ = -10 V), we obtain steeper behavior, close to $1/T_e^2$, which corresponds to $\delta = 3$.

\end{document}